\documentclass[12pt,pra,aps,amssymb,amsfonts,amsmath,tightenlines]{revtex4}
\usepackage{dsfont}
\usepackage[dvips]{graphicx}
\usepackage{amssymb,amsfonts}
\newcommand{\rd}{\mbox{$\rm d$}}

\begin{document}

\title{Classical fields as statistical states}

\author{Dorje~C.~Brody${}^1$ and Lane~P.~Hughston${}^2$}

\affiliation{Department of Mathematics, Imperial College London,
London SW7 2BZ, UK}

%\date{\today}

\begin{abstract}
We present a rough outline for an idea that characterises the
observed, macroscopic realisation of the electromagnetic field
in terms of a probability distribution on the underlying quantum
electrodynamic state space.
\end{abstract}

%\pacs{no pacs}

\maketitle

%\vskip .4cm

%{\bf Introduction}

In this note we sketch out some tentative thoughts on the
relation between microscopic and macroscopic states in quantum
theory. The idea is that classical fields (e.g., electromagnetism,
or weak-field gravity) should be thought of as {\it statistical}
states. Our starting point is to take the view that the physical
world is inherently quantum mechanical. The problem is thus not
how to quantise a given classical theory, but rather how to
classicalise the quantum theory appropriately
at large scales. Starting with a general multi-particle quantum
system characterised by a large state space and a myriad of
associated observables, our task is to specify those states of the
system that correspond, in some reasonable sense, to the classical
macroscopic configurations observed in practice.

This is an interesting question, because it ties in with some of the
major open issues in quantum theory that may be of relevance to
practical considerations. Even in the case of electromagnetism, it
has to be appreciated that there is no general agreement as to what
constitutes the precise relation between {\it microscopic} and
{\it macroscopic} realisations of the electromagnetic field, despite
the fact that classical and quantum electrodynamics are both well
developed theories. Sometimes it is suggested that the coherent
states of electromagnetism correspond to classical electrodynamic
fields, but the arguments supporting this idea are not entirely
convincing. Coherent states, to be sure, are in one-to-one
correspondence with classical solutions of Maxwell's equations---in
particular, to nonsingular, normalisable solutions. These states
also have the property that they saturate the uncertainty
lower bounds for measurements of the field operator. However,
there is no explanation for why this should be a `natural'
configuration for the electromagnetic state space. Since coherent
states are pure {\it quantum} states, we are left to wonder if
it is possible that these states could {\it remain} pure on a
macroscopic scale. This is questionable.

To put the matter another way, we expect a pure state to have low
entropy by any reasonable definition, whereas for the quasi-stable
nature of a classical field configuration, a high-entropy state
would be the more plausible candidate. Then we could invoke some
form of the second law of thermodynamics to explain the natural
occurrence of such configurations.

Now let us try to build up a model along these lines in more
precise terms. Suppose we consider a complex Hilbert space
${\cal H}$, for which we denote the associated multi-particle
bosonic Fock space ${\cal F}$. We use Greek indices for elements
of ${\cal H}$, and Roman indices for elements of ${\cal F}$. Thus, if
$\xi^{\alpha}\in {\cal H}$ and $\eta_{\alpha}\in {\cal H}^{*}$
(the dual space), then for their inner product we write
$\eta_{\alpha}\xi^{\alpha}$. Likewise, if $\Psi^{a}\in{\cal F}$ and
$\Phi_{a}\in{\cal F}^{*}$, then we can form the inner product
$\Phi_{a}\Psi^{a}$. An element $\Psi^{a}$ of ${\cal F}$ is given,
more explicitly, by a normalisable set of symmetric tensors in the
space
$({\bf C}, {\cal H},{\cal H}\otimes{\cal H},{\cal H}\otimes{\cal H}
\otimes{\cal H}, \cdots)$ given by
$\Psi^{a}=\{ \psi, \psi^{\alpha}, \psi^{(\alpha\beta)},
\psi^{(\alpha\beta\gamma)}, \cdots\}$ with the inner product
\begin{eqnarray}
\Phi_{a}\Psi^{a} = \phi\psi + \phi_{\alpha}\psi^{\alpha} +
\phi_{\alpha\beta}\psi^{\alpha\beta} + \phi_{\alpha\beta\gamma}
\psi^{\alpha\beta\gamma} + \cdots\ .
\end{eqnarray}
We have in mind, in particular, the case where ${\cal H}$ is the
Hilbert space of positive-frequency square-integrable solutions of
Maxwell's equations, equipped with the usual gauge independent
Hermitian inner product. Then ${\cal F}$ is the multi-particle
photon space of quantum electrodynamics, and the general element
of ${\cal F}$ determines a superposition of states consisting of
various numbers of photons, where the photon number is the rank of
the corresponding tensor. In addition we require a specification of
the electromagnetic field operators. The creation operator
$C^{\ a}_{\alpha b}$ is a map from ${\cal F}$ to
${\cal F}\otimes{\cal H}^{*}$ and the annihilation operator
$A^{\alpha a}_{\ b}$ is a map from ${\cal F}$ to ${\cal F}
\otimes{\cal H}$. They satisfy the commutation relations
$C^{\ a}_{\beta b}A^{\alpha b}_{\ c} - A^{\alpha a}_{\ b}
C^{\ b}_{\beta c} = \delta^{\alpha}_{\beta}\delta^{a}_{c}$. The
specific actions of $C$ and $A$ on ${\cal F}$ (see Geroch 1971)
are not required here.

The pure states of quantum electrodynamic systems are not
represented by elements of ${\cal F}$, but rather by points in the
associated {\it projective} Fock space ${\sl\Gamma}$. Let
$x$ denote a typical point in ${\sl\Gamma}$, and $\Psi^{a}(x)$ a
point in the fibre above $x$. Here, we think of ${\cal F}$ as a
fibre space over ${\sl\Gamma}$. Then the expectation of the
annihilation operator $A^{\alpha a}_{\ b}$, conditional to a
pure quantum state $x$, can be represented by a map
$A^{\alpha}(x)$ from ${\sl\Gamma}$ to ${\cal H}$, given by
\begin{eqnarray}
A^{\alpha}(x) = \frac{{\bar\Psi}_{a}(x)A^{\alpha a}_{\ b}
\Psi^{b}(x)}{{\bar\Psi}_{c}(x)\Psi^{c}(x)} .
\end{eqnarray}
Note that $A^{\alpha}(x)$ is independent of the scale of
$\Psi^{a}(x)$. Thus each quantum electrodynamic state is
associated with a unique classical field, given by the map
$x\in{\sl\Gamma}\rightarrow A^{\alpha}(x)\in {\cal H}$.

Now suppose we are given a classical solution
$\xi^{\alpha}\in{\cal H}$ of Maxwell's equations, and
we wish to construct a state on ${\sl\Gamma}$ to which
$\xi^{\alpha}$ should correspond in some natural {\it physical}
sense. The general state on the multi-particle photon state space
${\sl\Gamma}$ is given by a {\it probability distribution} $\rho(x)$
over ${\sl\Gamma}$. Thus, if $\rd x$ represents the natural
volume element on ${\sl\Gamma}$ associated with the Fubini-Study
metric, we have
\begin{eqnarray}
\int_{\sl\Gamma}\rho(x)\rd x = 1 .
\end{eqnarray}
The associated density matrix $\rho^{a}_{b}$, which contains
sufficient information to value the expectations of linear
observables, is given by
\begin{eqnarray}
\rho^{a}_{b} = \int_{\sl\Gamma} \rho(x)
\frac{{\bar\Psi}_{b}(x)\Psi^{a}(x)}
{{\bar\Psi}_{c}(x)\Psi^{c}(x)}\, \rd x  .
\end{eqnarray}
For example, the expectation of the annihilation operator can be
written in the form:
\begin{eqnarray}
\rho^{a}_{b}A^{\alpha b}_{\ a} = \int_{\sl\Gamma}
\rho(x) A^{\alpha}(x) \rd x .
\end{eqnarray}
We come to our key hypothesis. Our suggestion is that, for a
given classical field configuration $\xi^{a}$, the associated
physical quantum electrodynamic state is given by the probability
density function $\rho(x)$ that maximises the entropy function
\begin{eqnarray}
S_{\rho} = -\int_{\sl\Gamma}\rho(x)\ln\rho(x) \rd x
\end{eqnarray}
over ${\sl\Gamma}$, subject to the constraint
\begin{eqnarray}
\xi^{a} = \int_{\sl\Gamma}\rho(x)A^{\alpha}(x) \rd x .
\end{eqnarray}
A standard line of argument (cf. {\cite{2}) then shows that the
choice of $\rho(x)$ that maximises $S_{\rho}$ is given by a
{\it grand canonical ensemble} on ${\sl\Gamma}$ of the form:
\begin{eqnarray}
\rho(x) = \exp[-\mu_{\alpha}A^{\alpha}(x) -
{\bar\mu}^{\alpha}C_{\alpha}(x)]/Z(\mu) .
\end{eqnarray}
Here, the partition function $Z(\mu)$ is given by the integral
\begin{eqnarray}
Z(\mu) = \int_{\sl\Gamma}\exp\left[ -\mu_{\alpha}A^{\alpha}(x)
- {\bar\mu}^{\alpha}C_{\alpha}(x)\right] \rd x .
\end{eqnarray}
The `chemical potential' $\mu_{\alpha}\in{\cal H}^{*}$ arises as
a Lagrange multiplier in the variation analysis, and is determined
by the relation
\begin{eqnarray}
\frac{\partial\ln Z(\mu)}{\partial\mu_{\alpha}} = \xi^{\alpha}
\end{eqnarray}
for the given value of $\xi^{\alpha}$. Thus the dual field
$\mu_{\alpha}$ is {\it thermodynamically conjugate} to the
expectation of annihilation operator $\xi^{\alpha}$.

The manifold ${\sl\Gamma}$ is foliated by the level surfaces of
$A^{\alpha}(x)$, and the grand canonical distribution $\rho(x)$
is constant on each such surfaces. If we introduce the manifold
${\cal C}\in{\sl\Gamma}$ consisting of coherent states, then
${\sl\Gamma}$ can be viewed as a fibre space over ${\cal C}$,
because each level surface of the expectation of the annihilation
operator intersects ${\cal C}$ at one point, namely, the coherent
state corresponding to the given value of $A^{\alpha}(x)$. Coherent
states are those points $x$ for which the eigenvalue relation
$A^{\alpha a}_{\ b}\Psi^{b}(x)=\xi^{\alpha}\Psi^{a}(x)$ is
satisfied for some $\xi^{\alpha}$.

%%%%%%%%%%%%%%%%%%%%%%%%%%%
\begin{figure}
\begin{center}\vspace{-0.0cm}
  \includegraphics[scale=0.65]{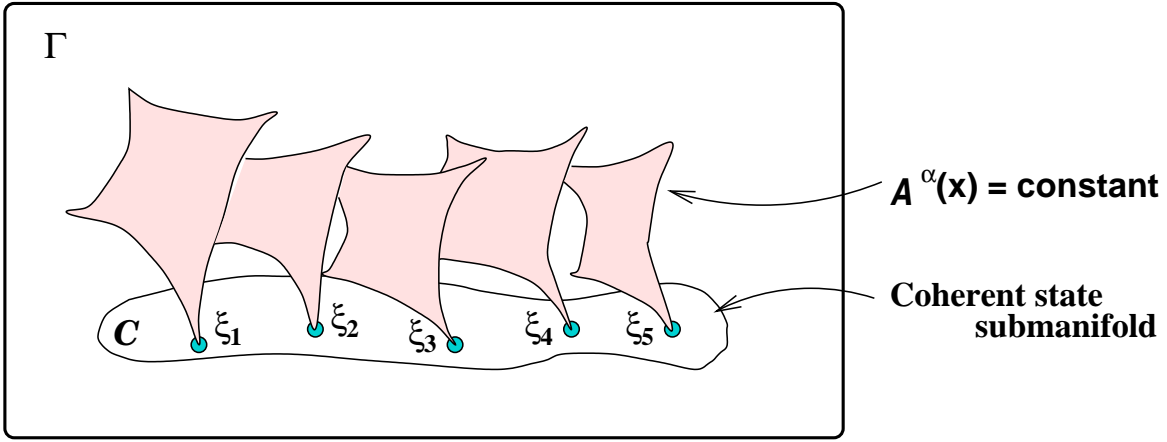}
  \vspace{-0.0cm}
  \caption{Foliation of projective Fock space ${\sl\Gamma}$ by level
surfaces of the expectation of the annihilation operator
$A^{\alpha}(x)$. Each such surface intersects the
manifold ${\cal C}$ of coherent states at a point.
  \label{fig:1}
  }
\end{center}
\end{figure}
%%%%%%%%%%%%%%%%%%%%%%%%%%%%

There are many interesting analogies that follow from the ideas
suggested above, and it is tempting to take them seriously. For
example, the phenomenon of classicalisation can be viewed as a
consequence of the second law of thermodynamics (cf. \cite{3}).
Suppose we have a large region of essentially classical
configurations (the laboratory) and a small region (the
experimental region) where we create, say, a region of pure
quantum state. Initially the experimental region is insulated from
the rest of the laboratory, but after the shield is removed the
experimental region `decoheres' by adjusting itself to the chemical
potential $\mu_{\alpha}$ of the laboratory. This follows from the
requirement of matching chemical potentials (in this case,
electromagnetic fields) for systems in equilibrium. More precisely,
the chemical potential of the laboratory has to shift
slightly to accommodate the experimental region, and as this happens
the pure state of the experimental region decoheres, or
classicalises, just in such a way as to match up with the chemical
potential of the laboratory, so that the combined system is now
characterised by a new classical state, say $\nu_{\alpha}$, which
differs from the original $\mu_{\alpha}$ by an essentially
negligible perturbation.

%%%%%%%%%%%%%%%%%%%%%%%%%%%
\begin{figure}
\begin{center}\vspace{-0.0cm}
  \includegraphics[scale=0.65]{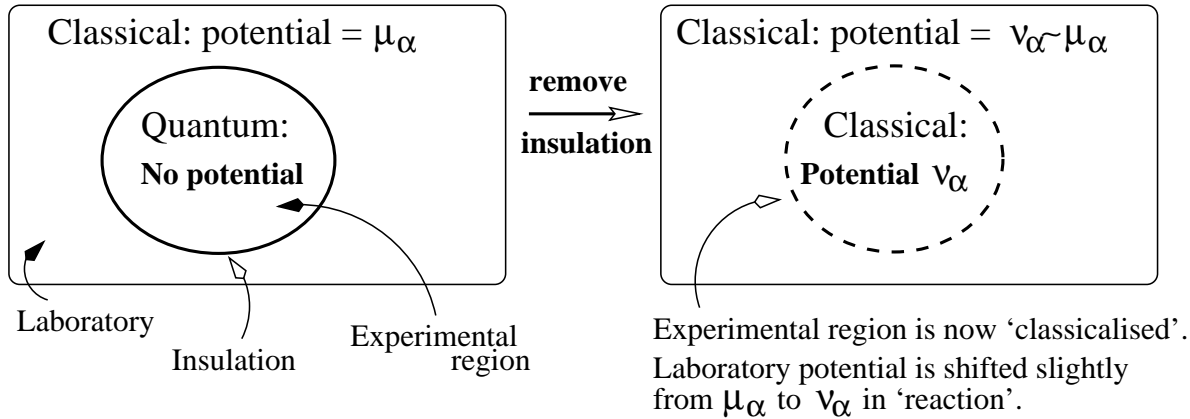}
  \vspace{-0.0cm}
  \caption{Classicalisation of a quantum field. An experimental
region confining a pure quantum field is insulated from the
environment of the surrounding laboratory, which is permeated
with a classical field of chemical potential $\mu_{\alpha}$. After
the insulation is removed the pure quantum state decoheres as
it comes into equilibrium with the environment, and acquires a
potential $\nu_{\alpha}$ equal to that of the laboratory,
which has shifted very slightly in response to its interaction
with the quantum field.
  \label{fig:2}
  }
\end{center}
\end{figure}
%%%%%%%%%%%%%%%%%%%%%%%%%%%%

In the idea outlined above, we implicitly assume that there exists
a dynamical mechanism leading to a kind of Boltzmann's $H$-theorem
for quantum electromagnetism. However, such a mechanism may not
exist as such in quantum physics, in which case the resulting
`equilibrium' (classical) distribution would have to be altered.
For example, we might be led to the density matrix
\begin{eqnarray}
\rho^{a}_{b} = \exp(-\mu_{\alpha}A^{\alpha a}_{\ b}
- {\bar\mu}^{\alpha}C^{\ a}_{\alpha b})/Q(\mu) ,
\end{eqnarray}
where the expectation $\rho^{a}_{b}A^{\alpha b}_{\ a}=\xi^{\alpha}$
determines $\mu_{\alpha}$ and $Q(\mu)$ is chosen so that
$\rho^{a}_{a}=1$. Nevertheless, the idea of representing classical
states as statistical distributions is legitimate, and it also ties
in naturally with the foundations of statistical mechanics, where the
aim is to account for the observed characteristics of macroscopic
physics in terms of an underlying microscopic dynamics. We hope to
exploit the idea further.

\vskip .4cm 

\begin{footnotesize}
\noindent Addresses when the paper was drafted: ${}^1$DAMTP, Silver Street, 
Cambridge CB3 9EW; ${}^2$Mathematics Department, King's College 
London, The Strand, London WC2R 2LS
\end{footnotesize}


\begin{enumerate}

\bibitem{1} R.~Geroch, Ann. Phys. {\bf 62}, 582 (1971).

\bibitem{2} D.~C.~Brody $\&$ L.~P.~Hughston, J. Math. Phys.
{\bf 39}, 6502 (1998); {\bf 40}, 12 (1999).

\bibitem{3} D.~C.~Brody $\&$ L.~P.~Hughston, Proc. Roy. Soc.
London A, {\bf 455}, 1683 (1999).

\end{enumerate}
\end{document}